\documentclass[12pt]{article}
\usepackage{epsf}
\usepackage{amsmath}
\usepackage{graphicx}
\usepackage{color}
\usepackage{cite}
\usepackage{bm}
\usepackage[colorlinks,citecolor=blue]{hyperref}
\renewcommand{\thefootnote}{\#\arabic{footnote}}

\newcommand{\gtrsim}{ \mathop{}_{\textstyle \sim}^{\textstyle >} }
\newcommand{\lesssim}{ \mathop{}_{\textstyle \sim}^{\textstyle <} }

\renewcommand{\thefootnote}{\fnsymbol{footnote}}
\def\thefootnote{\fnsymbol{footnote}}

\begin{document}

\begin{titlepage}

\begin{center}

\vskip .75in

{\Large \bf Blue-tilted inflationary tensor spectrum \vspace{3mm}  \\ and reheating  
in the light of NANOGrav results} \vspace{10mm}

\vskip .75in

{\large
Sachiko~Kuroyanagi$^{1,2}$,
Tomo~Takahashi$^{3}$ and 
Shuichiro~Yokoyama$^{4,5}$
}

\vskip .45in

{\em
$^{1}$Instituto de F\'isica Te\'orica UAM-CSIC, Universidad Aut\'onoma de Madrid, \\Cantoblanco, 28049 Madrid, Spain  \vspace{3mm} \\
$^{2}$Department of Physics, Nagoya University, Nagoya, Aichi 464-8602, Japan  \vspace{3mm} \\
$^{3}$Department of Physics, Saga University, Saga 840-8502, Japan \vspace{3mm} \\
$^{4}$Kobayashi Maskawa Institute, Nagoya University, Aichi 464-8602, Japan  \vspace{3mm} \\
$^{5}$Kavli Institute for the Physics and Mathematics of the Universe (WPI), \\
Todai institute for Advanced Study, University of Tokyo, Kashiwa, Chiba 277-8568, Japan

}

\end{center}
\vskip .5in

\begin{abstract}

We discuss the possibility of explaining the recent NANOGrav results by inflationary gravitational waves (IGWs) with a blue-tilted primordial spectrum. Although such IGWs can account for the NANOGrav signal without contradicting the upper bound on the tensor-to-scalar ratio at the cosmic microwave background scale, the predicted spectrum is in strong tension with the upper bound on the amplitude of the stochastic gravitational wave background by big-bang nucleosynthesis (BBN) and the second LIGO-Virgo observation run.  However, the thermal history of the Universe, such as reheating and late-time entropy production, affects the spectral shape of IGWs at high frequencies and permits evading the upper bounds. We show that, for the standard reheating scenario, when the reheating temperature is relatively low, a blue tensor spectrum can explain the recent NANOGrav signal without contradicting the BBN and the LIGO-Virgo constraints. We further find that, when one considers a late-time entropy production, the NANOGrav signal can be explained even for an instant reheating scenario.

\end{abstract}

\end{titlepage}

\renewcommand{\thepage}{\arabic{page}}
\setcounter{page}{1}
\renewcommand{\thefootnote}{\#\arabic{footnote}}
\setcounter{footnote}{0}

\section{Introduction \label{sec:intro}}

Stochastic gravitational wave background (SGWB) can be used as a powerful probe of the Universe. Many possible sources of SGWB, both from cosmological and astrophysical processes, have been suggested and investigated rigorously (see, e.g., \cite{Kuroyanagi:2018csn} and references therein). If a SGWB was detected, it would provide important implications for cosmology and astrophysics.

Recently, the North American Nanohertz Observatory for Gravitational Waves (NANOGrav) has reported a possible detection of a SGWB with their 12.5 year data set of the pulsar timing array \cite{Arzoumanian:2020vkk}. Although one needs a confirmation of quadrupolar spatial correlations to claim the detection of a SGWB \cite{Pol:2020igl}, the NANOGrav results have already stimulated a lot of work to pursue a possible source of the cosmological signal \cite{Blasi:2020mfx,Ellis:2020ena,Vaskonen:2020lbd,DeLuca:2020agl, Nakai:2020oit, Buchmuller:2020lbh, Addazi:2020zcj, Ratzinger:2020koh, Kohri:2020qqd, Samanta:2020cdk, Vagnozzi:2020gtf, Neronov:2020qrl, Bian:2020bps, Namba:2020kij, Li:2020cjj, Sugiyama:2020roc, Domenech:2020ers, Bhattacharya:2020lhc, Abe:2020sqb, Kitajima:2020rpm,1827858,1827882}. 

Among various sources of SGWB, the inflationary gravitational waves (IGWs) are one of the most interesting sources to explore as a possible explanation of the recent NANOGrav results. In the inflationary Universe, not only the scalar fluctuations, which source the cosmic large-scale structure, but also the tensor fluctuations, which could be observed as a SGWB, are generated from quantum fluctuations of spacetime. The IGWs naturally have a broad power spectrum, whose amplitude depends on the inflationary model. Since the IGWs on large scales (corresponding the frequency of $\sim 10^{-16}$Hz) can be well constrained by  cosmic microwave background (CMB), one needs a blue-tilted tensor spectrum to explain the NANOGrav results at $\sim 10^{-8}$Hz without contradicting the upper bound on the tensor-to-scalar ratio obtained by recent CMB experiments, such as the Planck satellite. Although it is generally difficult to realize such a blue-tilted tensor spectrum in the standard slow-roll inflation in general relativity, possible models have been proposed and discussed in the context of modified theories of gravity ({\it e.g.} through the Galileon-type interaction term or the Gauss-Bonnet coupling) or non-standard inflation models ({\it e.g.} solid inflation) \cite{Piao:2004tq, Gruzinov:2004ty, Satoh:2008ck, Kobayashi:2010cm, Endlich:2012pz, Koh:2014bka, Cannone:2014uqa, Cai:2014uka, Cai:2015yza, Cai:2016ldn, Ricciardone:2016lym, Koh:2018qcy, Fujita:2018ehq, Iacconi:2019vgc, Mishima:2019vlh, Iacconi:2020yxn, 1827882}.

When considering a blue-tilted tensor spectrum, we also need to check whether the spectral amplitude at high frequencies (above $\sim 10^{-10}$Hz) is consistent with the big-bang nucleosynthesis (BBN) bound on an extra radiation component. Furthermore, the LIGO-Virgo interferometer network currently provides a tighter upper bound on the SGWB amplitude at the frequency of $\sim 100~{\rm Hz}$.\footnote{
Scalar fluctuations can be induced by IGWs through non-linear (second-order) effects, which can lead to the overproduction of primordial black holes (PBHs). Thus, the current upper bounds on the abundance of PBHs can also provide comparable constraints on the amplitude of the IGWs at relatively higher frequency region \cite{Nakama:2015nea, Nakama:2016enz}.
}  
In fact, the constraints on the SGWB with a blue-tilted spectrum by BBN and LIGO-Virgo are quite strong and hence it is nontrivial to determine whether IGWs can explain the recent NANOGrav signal without conflicting with the BBN and LIGO-Virgo constraints, which is the issue we will investigate in this paper. 

As the simplest thermal history of the Universe, one can consider the case where reheating occurred instantaneously just after inflation.  In this case, a blue-tilted spectrum for explaining the NANOGrav result is ruled out by the BBN constraint, as has been pointed out in \cite{Vagnozzi:2020gtf,Bhattacharya:2020lhc}.\footnote{Although the first version of ref.~\cite{Vagnozzi:2020gtf} did not mention much about the reheating temperature $T_R$ has been added in the second version which appeared on the arXiv after our paper.} However, in the conventional scenario of reheating, the oscillating inflaton-dominated phase, where the expansion of the background Universe behaves like that in a matter-dominated (MD) era, lasts for a certain period of time after inflation, followed by the radiation-dominated (RD) Universe. It has been shown that the power spectrum of the IGWs which re-entered the horizon during a MD phase get suppressed by a frequency dependence of $\propto f^{-2}$~\cite{Seto:2003kc,Boyle:2005se, Nakayama:2008wy,Kuroyanagi:2008ye,Kuroyanagi:2010mm,Kuroyanagi:2011fy,Kuroyanagi:2014qza}, while the modes which entered the horizon during the RD phase have a dependence $\propto f^0$. Thus, by considering an early MD phase after inflation, the GW amplitude at high frequencies can be reduced, which enables us to evade the BBN and LIGO-Virgo constraints. Therefore, the model of reheating significantly affects constraints on the tensor spectral index $n_t$ by BBN and LIGO-Virgo~\cite{Kuroyanagi:2014nba}. Indeed, as we will see in this paper, for example if we assume the tensor-to-scalar ratio $r=0.06$ at the CMB scale, a blue-tilted spectrum can explain the NANOGrav signal without contradicting the upper bounds from BBN and LIGO-Virgo, when the reheating temperature is lower than $T \simeq 10^3\,{\rm GeV}$.

Furthermore, in some scenarios of the early Universe, late-time entropy production can occur due to the existence of a scalar field such as a modulus. In this case, an early MD phase is realized after the RD period followed by reheating due to the inflaton. Namely, in this scenario, MD phases appear twice between the end of inflation and the BBN epoch, which accordingly makes the GW spectra different from the case of the standard thermal history~\cite{Nakayama:2009ce,Kuroyanagi:2013ns, Li:2016mmc, DEramo:2019tit}. As mentioned above, the GW spectrum gets suppressed for the modes which entered the horizon during MD phase, and thus the constraints from BBN and LIGO-Virgo can be alleviated by late-time entropy production as well~\cite{Kuroyanagi:2014nba}. We will see that, if the duration of the late-time entropy production is long enough, even IGWs with instant reheating at $T_R \sim 10^{15}\,{\rm GeV}$ can explain the NANOGrav signal while satisfying the constraints from BBN and LIGO-Virgo.

The structure of this paper is the following. In the next section, we briefly summarize the formulas to calculate the spectrum of IGWs and how IGWs with a blue-tilted spectrum are constrained by the BBN and the LIGO-Virgo upper bounds. In Section~\ref{sec:nanograv}, we discuss the possibility of explaining the NANOGrav result with IGWs with a blue-tilted spectrum by taking into account reheating and late-time entropy production, and then explore the parameter space allowed by both BBN and LIGO-Virgo. The last section is devoted to conclusion.

\section{The spectrum of IGWs and upper bounds by BBN and LIGO-Virgo\label{sec:spectrum}}

Here we briefly summarize the formulas to calculate the spectrum of IGWs and upper bounds on the amplitude of the SGWB by BBN and the LIGO-Virgo interferometer network. For details, see Ref.~\cite{Kuroyanagi:2014nba} and the references therein.

\subsection{The spectrum of IGWs}
GWs are described as tensor metric perturbations $h_{ij} ( \tau, \bm{x})$ in the flat Friedmann-Lema\^{i}tre-Robertson-Walker background:
\begin{equation}
ds^2 = a(\tau) \left[ -d\tau^2 + \left( \delta_{ij} + h_{ij} (\tau, \bm{x} ) \right) \, dx^i dx^j  \right] \,, 
\end{equation}
where $\tau$ is the conformal time, $a(\tau)$ is the scale factor, and $h_{ij}$ satisfies the transverse-traceless condition, $\partial^i h_{ij} = h^i_{~i} = 0$. Using the Fourier transformation,
\begin{equation}
h_{ij} (\tau, {\bm x}) = \sum_\lambda \int \frac{dk^3}{(2\pi)^{3/2}} \, e^{i \bm x \cdot  \bm k} \, \epsilon_{ij}^\lambda ({\bm k}) \, h_{\bm k}^\lambda (\tau) \,,
\end{equation}
with $\epsilon^\lambda_{ij} ({\bm k})$ being the polarization tensor, the energy density of GWs can be written as 
\begin{equation}
\rho_{\rm GW} = \frac{1}{32 \pi G} \int d \ln k \, \left( \frac{k}{a} \right)^2 \, \frac{k^3}{\pi^2} \, \sum_\lambda \, \left| h_{\bm k}^\lambda \right|^2 \,.
\end{equation}
By defining the power spectrum of primordial tensor perturbations
generated during inflation as 
\begin{equation}
{\cal P}_{T, {\rm prim}} (k) = \frac{k^3}{\pi^2} \sum_\lambda |h_{\bm k, \ast}^\lambda |^2 \,,
\end{equation}
where $*$ denotes the value at the end of inflation,\footnote{
Here, for simplicity, we assume that the tensor perturbations on super-horizon scales stay constant in time after the end of inflation.
In case where the super-horizon tensor perturbations evolve even after the end of inflation, we need to modify the transfer function, $T_T(k)$, which will be given by Eqs.~(\ref{eq:transfer}) and \eqref{eq:transfer_F}.
}
one can write down the GW energy density parameter today as 
\begin{equation}
\Omega_{\rm GW} 
\equiv 
 \frac{1}{\rho_{\rm crit}} \frac{ d \rho_{\rm GW} }{d \ln k} 
 = \frac{1}{12} \left( \frac{k}{a_0 H_0} \right)^2 \, T_T^2(k) \, {\cal P}_{T, {\rm prim}} (k) \,,
\end{equation}
where $\rho_{\rm crit} = 3 H_0^2 /(8\pi G)$ is the critical density of the universe and $T_T(k)$ is the transfer function, which is defined as $T^2_T (k) =  |h_{\bm k, 0}^\lambda |^2/ |h_{\bm k, \ast}^\lambda |^2$ with $0$ denoting the value at the present time.

The primordial tensor spectrum, ${\cal P}_{T, {\rm prim}} (k)$, is often parameterized as 
\begin{equation}
\label{eq:P_T}
{\cal P}_{T, {\rm prim}} (k) = A_T \left( \frac{k}{k_{\rm ref}} \right)^{n_T} \,,
\end{equation}
where $A_T $ is the amplitude at a reference scale $k_{\rm ref}$  and $n_T$ is the tensor spectral index.
The amplitude of the primordial tensor perturbations at the CMB scale is commonly characterized by the tensor-to-scalar ratio $r$ as
\begin{equation}
\label{eq:tensor-to-scalar }
r \equiv \frac{{\cal P}_{T, {\rm prim}} (k_{\rm ref})}{{\cal P}_{\zeta} (k_{\rm ref}) }  \,,
\end{equation}
where $ {\cal P}_{\zeta} (k)$ is the primordial power spectrum of  scalar perturbations. 
The spectral index $n_T$ can also be a scale-dependent variable and the dependence 
strongly relies on the model.\footnote{
The scale-dependence can be expressed by adding the higher-order terms of $\ln (k/k_{\rm ref})$ ({\it i.e.} the spectral running and the running of running), perturbatively.
For example, higher-order correction to the scale dependence for a standard slow-roll inflation is investigated in \cite{Kuroyanagi:2011iw}. For non-standard inflation models and alternative models of inflation, see, e.g. \cite{calcagni2020stochastic}.
}
Since this paper aims to perform a model-independent study, we simply assume it to be constant.

For the transfer function $T_T(k)$, we use the following fitting formulas. In the case of the standard reheating scenario, the universe experiences a MD (inflation $\phi$ oscillation-dominated) phase after inflation and the RD phase follows. In this case, the transfer function is given by \cite{Turner:1993vb,Nakayama:2008wy,Kuroyanagi:2014nba}:
\begin{eqnarray}
\label{eq:transfer}
T^2_T (k) &=& \Omega_m^2 \left( \frac{g_\ast (T_{k, \rm in})}{g_{\ast 0} } \right) \left( \frac{g_{\ast 0} }{g_{\ast s} (T_{k, \rm in})} \right)^{4/3} \left( \frac{3 j_1 (k\tau_0) }{k \tau_0} \right)^2 T^2_1 (x_{\rm eq}) T^2_2 (x_R)   \\ 
&&\hspace{80mm}  \textrm{(Standard reheating scenario)}\,,  \notag
\end{eqnarray}
where $\Omega_m$ is the present day energy density parameter for matter, $T_{k, \rm in}$ denotes the temperature at the time when mode $k$ enters the horizon, and  $g_\ast (T) $ and $g_{*s} (T) $ are the relativistic degrees of freedom and its counterpart for entropy, respectively. The subscript $0$ denotes the value at the present time, and we use $g_{*0}=3.36$ and $g_{*s0}=3.91$.  In the limit of $k\tau_0 \rightarrow 0$, the spherical Bessel function can be approximated as $j_1 (k\tau_0)\approx 1/\sqrt{2}k\tau_0$. The changes of $g_\ast (T) $ and $g_{*s} (T) $ affect the shape of the spectrum of IGWs~\cite{Watanabe:2006qe,Kuroyanagi:2008ye,Saikawa:2018rcs} and we use the fitting formula proposed in Ref.~\cite{Kuroyanagi:2014nba},
\begin{equation}
\label{eq:g_ast}
g_\ast (T_{k, \rm in}(k)) 
= g_{\ast 0} 
\left( \frac{A + \tanh \left[ -2.5 \log_{10} \left( \frac{k/(2\pi) }{2.5 \times 10^{-12} \, {\rm Hz}} \right) \right]}{ A+1} \right)
\left( \frac{B + \tanh \left[ -2.0 \log_{10} \left( \frac{k/(2\pi) }{6.0 \times 10^{-9} \, {\rm Hz}} \right) \right]}{B+1} \right) \,,
\end{equation}
where
\begin{eqnarray}
A &=& \frac{-1-  10.75/ g_{\ast 0}  }{-1+ 10.75/ g_{\ast 0}} \,,
\qquad
B = \frac{-1-  g_{\rm max}/ 10.75  }{-1+ g_{\rm max}/ 10.75 } \,.
\end{eqnarray}
Here $g_{\rm max}$ is the maximum value for $g_\ast$ and we take the sum of the standard model particles $g_{\rm max}=106.75$. The entropic counterpart $g_{*s} (T_{k, {\rm in}})$ can be obtained by replacing $g_{\ast 0}$ with $g_{*s 0}$ in Eq.~\eqref{eq:g_ast}. The function $T_1(x)$ and $T_2(x)$ are the fitting formulas corresponding to the changes of the spectral shape due to the radiation-matter equality and reheating, and respectively given by 
\begin{eqnarray}
T_1^2 (x) & = &  1 + 1.57 x + 3.42 x^2  \,, \\ [8pt]
T_2^2 (x) & = &  \left( 1 -0.22 x^{1.5} + 0.65 x^2 \right)^{-1} \,,
\end{eqnarray}
where $x_{\rm eq} \equiv k/k_{\rm eq}$ and $x_R \equiv k/k_R$ with 
\begin{eqnarray}
k_{\rm eq} &=& 7.1 \times 10^{-2} \, \Omega_m h^2 \, {\rm Mpc}^{-1} \,, \\ [8pt]
k_R &=& 1.7 \times 10^{14} \, \left( \frac{g_{\ast s} (T_R) }{106.75} \right)^{1/6} \left( \frac{T_R}{10^7 ~{\rm GeV}} \right)  \, {\rm Mpc}^{-1} \,.
\end{eqnarray}
Here each wavenumber corresponds to the mode which enters the horizon at the radiation-matter equality and the transition from the early MD phase to the RD phase, respectively, and $T_R$ is the so-called reheating temperature, which is the temperature of the Universe when reheating completes and the RD phase starts. 

In the scenario where the late-time entropy production occurs after reheating through the decay of an oscillating scalar field $\sigma$, the background evolution proceeds as MD (inflation $\phi$ oscillation-dominated) $\rightarrow$ RD $\rightarrow$ MD ($\sigma$-oscillation dominated) $\rightarrow$ RD. In such a case, the transfer function is given by \cite{Nakayama:2009ce,Kuroyanagi:2014nba}
\begin{eqnarray}
\label{eq:transfer_F}
T^2_T (k) &=& 
\Omega_m^2 \left( \frac{g_\ast (T_{k, \rm in})}{g_{\ast 0} } \right) \left( \frac{g_{\ast 0} }{g_{\ast s} (T_{k, \rm in})} \right)^{4/3} \left( \frac{3 j_1 (k\tau_0) }{k \tau_0} \right)^2 T^2_1 (x_{\rm eq}) T^2_2 (x_\sigma) T_3^2 (x_{\sigma R}) T_2^2 (x_{R'})  
\notag \\  \\
&& \hspace{70mm}   \textrm{(Late-time entropy production scenario)} \,, \notag
\end{eqnarray}
where $T_3^2(x)$ is the transfer function corresponding to the transition from the first RD phase to the  $\sigma$-oscillation dominated phase and given by 
\begin{equation}
T_3^2(x) = 1 + 0.59 x + 0.65 x^2.
\end{equation}
Here we define $x_\sigma \equiv k/ k_\sigma$ where $k_\sigma$ corresponds to the mode entering the horizon at the time when $\sigma$ decays into radiation, i.e., the second reheating, and is given by
\begin{eqnarray}
k_\sigma &=&  1.7 \times 10^{14}  \left( \frac{g_{\ast s} (T_\sigma) }{106.75} \right)^{1/6} \left( \frac{T_\sigma}{10^7~{\rm GeV}} \right) ~{\rm Mpc}^{-1},
\end{eqnarray}
with $T_\sigma$ being the temperature of the Universe when the decay of $\sigma$ field completes.  We also define $x_{\sigma R} = k/k_{\sigma R}$ and $x_{R'} = k/k_{R'}$, which are respectively given by
\begin{equation}
k_{\sigma R} = k_\sigma F^{2/3}, 
\qquad 
k_{R'} = k_R F^{-1/3}\,,
\end{equation}
where $F$ characterizes the amount of entropy produced by the decay of $\sigma$ and can be written as\footnote{
In the case of entropy production by an oscillating $\sigma$ field, 
$F$ can be given by \cite{Nakayama:2009ce}
\begin{equation*}
    F =\frac{\sigma_i^2}{6M_{\rm pl}^2} \frac{T_R}{T_\sigma}~,
\end{equation*}
with $\sigma_i$ being the initial amplitude of the $\sigma$ field.
}
\begin{equation}
\label{eq:F}
F \equiv \frac{s(T_\sigma) a^3 (T_\sigma)}{s(T_R) a^3 (T_R)} \,,
\end{equation}
with $s(T)$ being the entropy density. Here, $k_{\sigma R}$ corresponds to the mode entering the horizon when oscillating $\sigma$ begins to dominate the Universe. For later discussion, we also define the temperature corresponding to $k_{\sigma R}$ as $T_{\sigma R}$. Note that the wavenumber corresponding to the end of first reheating is modified by the factor of $F^{-1/3}$ compared to the case without second reheating even for the same reheating temperature. We use the subscript $R'$ instead of $R$ in order to distinguish them.

By using the fitting formulas for the transfer functions given here, one can calculate the spectra of IGWs for scenarios with the standard reheating and late-time entropy production. Throughout the paper, we use the Planck 2018 cosmological parameters for a flat $\Lambda$-CDM universe; $h=0.6737$, $\Omega_m =0.3147$, $\Omega_\Lambda=1-\Omega_m$, and $\ln ({\cal P}_{\zeta}10^{10}) = 3.043$ at $k_{\rm ref} = 0.05~{\rm Mpc}^{-1}$ \cite{Aghanim:2018eyx}. The upper bound on the tensor-to-scalar ratio has been updated in the literature~\cite{Akrami:2018odb,Ade:2018gkx,Tristram:2020wbi}. In the following, we use $r=0.06$ at $k_{\rm ref} =0.05\,{\rm Mpc}^{-1}$ as a reference upper bound for $r$ on large scales \cite{Ade:2018gkx}.

\subsection{BBN constraint}

Primordial GWs contribute to the total energy density of extra relativistic species during BBN. In order not to spoil the BBN through the expansion rate of the Universe, 
\footnote{
Although CMB can also put a constraint on an extra radiation component, in most analysis of CMB, the extra radiation component is implemented as a fluid similar to neutrino. However, fluctuations of the extra radiation originated from GWs behave differently from those of such a fluid and the evolution of fluctuations would be different, and hence we do not consider the constraint from CMB in this paper.
} 
the energy density of the SGWB should satisfy
\begin{equation}
\label{eq:BBN_Neff}
\int_{f_1}^{f_2} d (\ln f) \Omega_{\rm GW} (f) h^2 \le 5.6 \times 10^{-6} \left( N_{\rm eff}^{\rm (upper)} - 3.046 \right) \,,
\end{equation}
where $N_{\rm eff}^{\rm (upper)}$ is the upper bound on the effective number of relativistic degrees of freedom. To evaluate the constraint from BBN, we adopt the 2$\sigma$ upper limit for $N_{\rm eff}$ from BBN (observations of $^4$He  and D) as $N_{\rm eff}^{\rm (upper)} = 3.41$ \cite{Cyburt:2015mya}.
In this paper, we take the lowest reheating temperature to be $T_R=10^{-2}$~GeV. 
Although, in a scenario with low-reheating temperature, neutrinos may not have enough time to be thermalized before their decoupling, the reheating temperature of $T_R=10^{-2}$~GeV still allows them to be thermalized and realize $N_{\rm eff} = 3.046$ \cite{Kawasaki:2000en, Hannestad:2004px}. Therefore we can safely use the bound Eq.~\eqref{eq:BBN_Neff}.

The lower limit of the integral in the left hand side is the frequency of the mode entering the horizon at the BBN epoch and we take $f_1 = 10^{-10}~{\rm Hz}$. The upper limit of the integral $f_2$ corresponds to the highest frequency of GWs, which is determined by the Hubble expansion rate at the end of inflation, and given by
\begin{equation}
f_2 = \frac{a_{\rm end}H_{\rm end}}{2\pi} \approx 1.1\times 10^8
\left(\frac{T_R}{10^{15}{\rm GeV}}\right)^{1/3}\left(\frac{H_{\rm end}}{10^{14}{\rm GeV}}\right)^{1/3} F^{-1/3}~{\rm Hz}
\,.
\end{equation}
Typically, when we consider the MD phase during reheating, the high-frequency IGWs are suppressed and do not contribute to the integral. Thus, although the value of $f_2$ has small dependencies on $T_R$, $H_{\rm end}$ and $F$, they do not affect the BBN constraint unless we consider instant reheating.

\subsection{LIGO-Virgo constraint}

Non-detection of a SGWB by the Advanced LIGO-Virgo detector network in the first and second observing run (O1 ans O2) has provided an upper bound on the amplitude of the SGWB as \cite{LIGOScientific:2019vic}
\begin{equation}
\Omega_{\rm GW} \le 6.0 \times 10^{-8},
\end{equation}
at the reference frequency of $f_{\rm LV}=25$Hz for a frequency-independent (flat) background. The frequency band containing  99\% of the sensitivity is $f= 20 - 81.9 \, {\rm Hz}$ for O2.

Note that this is an upper limit obtained for $\alpha=0$, where $\alpha$ is a parameter to characterize the spectral tilt at the LIGO frequency, $\Omega_{\rm GW}= \Omega_{\rm GW,\alpha} (f/f_{\rm LV})^\alpha$. The upper limit changes for different values of $\alpha$. In order to take this into account, we use the following formula adding the theoretically obtained $\alpha$ dependence (for details, see \cite{Kuroyanagi:2014nba}), 
\begin{equation}
\label{eq:LIGO_Virgo}
\Omega_{\rm GW} \le 6.0 \times 10^{-8}\sqrt{\frac{5-2\alpha}{5}}
\left(\frac{f_{\rm min}}{f_{\rm LV}}\right)^{-\alpha},
\end{equation}
where we take $f_{\rm min}=20$Hz. 

\section{Implications of the NANOGrav results for IGWs \label{sec:nanograv}}

Now we discuss implications of the recent NANOGrav results for IGWs with a blue-tilted spectrum. We calculate the spectra of IGWs and constraints from BBN and LIGO-Virgo by using the procedure presented in the previous section for the cases with the standard reheating and late-time entropy production scenario. Model parameters for the standard reheating scenario are: the amplitude of the IGWs, which we characterize by the tensor-to-scalar ratio $r$, the spectral index $n_T$, and the reheating temperature $T_R$. For the case of late-time entropy production scenario, in addition to $r, n_T$, and $T_R$, we also need to specify the second reheating temperature $T_\sigma$ and the amount of entropy produced by the decay of the $\sigma$ field, which is characterized by $F$.

\begin{figure}
\begin{center}
\includegraphics[width=\textwidth]{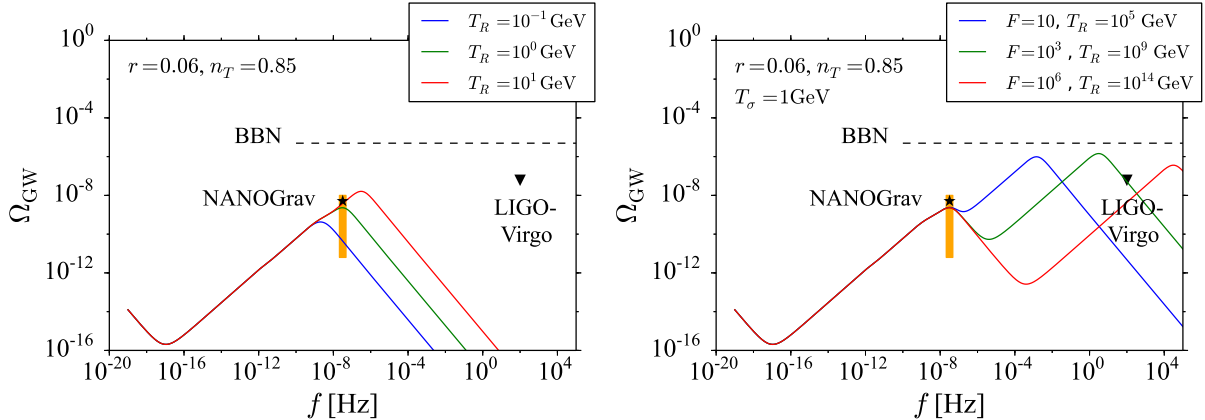}
\caption{Examples of spectra of IGWs for the scenarios with the standard reheating (left) and late-time entropy production (right). The amplitude of the SGWB indicated by the recent NANOGrav results (orange colored region) and the current upper bounds by BBN (black dashed line) and LIGO-Virgo (black triangle) are also shown. Note that the NANOGrav results give $A_{CP}=1.92\times 10^{-15}$ (median value) for a $f^{-2/3}$ power-law spectrum (black star), while a larger range of $A_{CP}$ is allowed if we consider different spectral tilt. \label{fig:GW_spectra} }
\end{center}
\end{figure}

In Fig.~\ref{fig:GW_spectra}, the spectra of IGWs for some parameter sets are shown. In the left panel, we show the spectra for the standard reheating scenario, assuming $r=0.06$ and $n_T =0.85$, for different reheating temperatures $T_R = 10^{-1},1,10$~GeV. In the right panel, we show the case of late-time entropy production, assuming $r=0.06$, $n_T =0.85$, $T_\sigma =1$~GeV and $F =10, 10^3, 10^6$ with $T_R = 10^5,10^9,10^{14}$~GeV, respectively. We also show the amplitude of the SGWB indicated by the recent NANOGrav result (orange colored region), and upper bounds by BBN (black dashed line) and LIGO-Virgo (black triangle).

In the following, we discuss whether IGWs with a blue-tilted spectrum can account for the recent NANOGrav signal without conflicting the constraints from BBN and LIGO-Virgo and investigate the parameter spaces.

\subsection{Case with standard reheating \label{sec:standard}}
\begin{figure}
\begin{center}
\includegraphics[width=\textwidth]{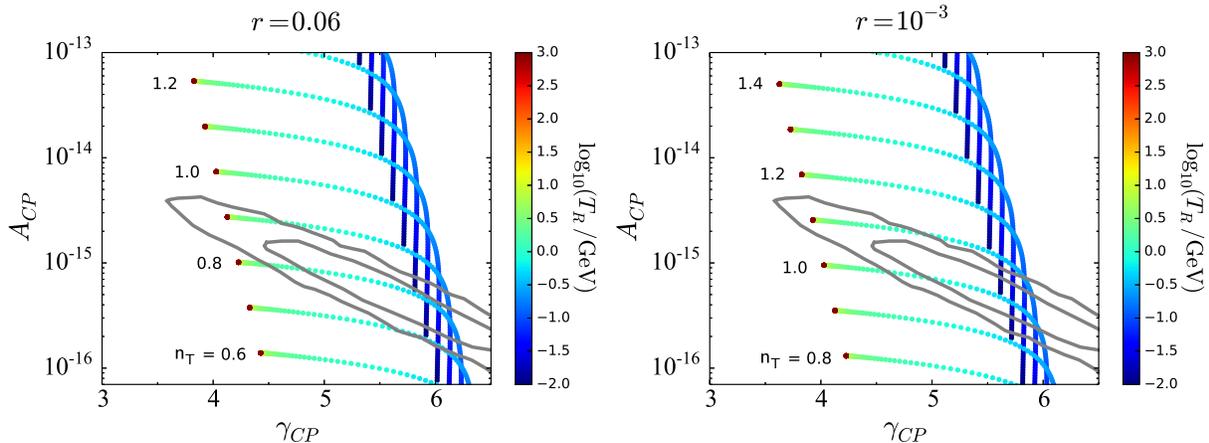}
\caption{Predictions of the IGWs with a blue-tilted spectrum for the standard reheating scenario. We scan over the spectral tilt $n_T$ and reheating temperature $T_R$ projected into the $\gamma_{CP}-A_{CP}$ plane, where $A_{CP}$ is the characteristic GW strain amplitude at $f=1{\rm yr}^{-1}$ and $\gamma_{CP}$ is the spectral index of the pulsar timing-residual cross-power spectral density. We fix the value of the tensor-to-scalar ratio at $r=0.06$ in the left panel and $r=10^{-3}$ in the right panel. The gray curves are the 1$\sigma$ and 2$\sigma$ posterior contours obtained by NANOGrav based on the five-frequency power-law fit \cite{Arzoumanian:2020vkk}. \label{fig:gamma_A} }
\end{center}
\end{figure}

We start with the case of the standard reheating scenario. The NANOGrav results have been reported in terms of $h_c(f)$, the power spectrum of the characteristic GW strain which can be related to $\Omega_{\rm GW}$ as $\Omega_{\rm GW}(f)=(2\pi^2/3H_0^2)f^2 h_c^2(f)$, and approximated as a power-law form:
\begin{equation}
h_c (f) = A_{\rm CP} \left( \frac{f}{f_{\rm yr}} \right)^{(3- \gamma_{\rm CP})/2}\,,
\end{equation}
where $A_{\rm CP}$ refers to the amplitude at the frequency $f = f_{\rm yr} = 1 ~{\rm yr}^{-1}$. When the spectral tilt at the pulsar timing scale is the same as the one at the CMB scale, power-law index $\gamma_{\rm CP}$ can be related to $n_T$ as $\gamma_{\rm CP} = 5 - n_T$. On the other hand, the value of $\gamma_{\rm CP}$ deviates from $5 - n_T$ if the spectral shape is altered due to the change of Hubble expansion rate during reheating and late-time entropy production.\footnote{
Note that the effect of $g_*$ change is seen around the pulsar timing frequencies, not at $f\sim f_{\rm yr}$ but around $f\sim 0.1~{\rm yr}^{-1}$ \cite{Kuroyanagi:2008ye}. This could slightly change the value of $\gamma_{CP}$ when we consider broad frequency range, but in our analysis, we estimate the spectral tilt at $f=f_{\rm yr}$ and thus the $g_*$ change does not affect the value of $\gamma_{\rm CP}$.}

In Fig.~\ref{fig:gamma_A}, we show the parameter scan in the $\gamma_{\rm CP}-A_{\rm CP}$ plane shown with the 1$\sigma$ and 2$\sigma$ posterior contours of the NANOGrav results obtained from the five-frequency power-law fit \cite{Arzoumanian:2020vkk}. In the left and right panels, the tensor-to-scalar ratio is fixed as $r=0.06$ and $10^{-3}$, respectively. The predictions for $A_{\rm CP}$ and $\gamma_{\rm CP}$ are shown for several values of $n_T$, corresponding to the curves from the bottom (low $n_T$) to the top (high $n_T$), and different values of reheating temperature $T_R$, represented by different colors. 

As expected, the amplitude of IGWs at the pulsar timing frequency $A_{\rm CP}$ gets larger for increasing $n_T$. We see from the colors that the dots degenerate at the same place when $T_R \gtrsim 10$~GeV, because in this case, the effect of reheating appears at frequencies higher than $f_{\rm yr}$ and the spectrum at the pulsar timing frequency stays the same. On the other hand, when $T_R \lesssim 10$~GeV, the value of $\gamma_{\rm CP}$ gets larger for smaller reheating temperatures since the spectrum becomes red-tilted at the pulsar timing frequency because of the bending in the spectrum induced by reheating. We see that the spectrum gets flattened (corresponding to $\gamma_{\rm CP} = 5$) when $T_R\sim 1$~GeV, and the curve with constant $n_T$ converges to the line of $\gamma_{\rm CP}= 7 - n_T$ for smaller $T_R$, since reheating induces the extra frequency dependence of $f^{-2}$. 
As we have mentioned, we take the lowest reheating temperature to be $T_R=10^{-2}$~GeV.

As seen from the figure, for the case of $r=0.06$, the NANOGrav signal can be IGWs for a broad range of reheating temperatures when the tensor spectral index is $n_T \sim 0.8$.  Even when $n_T \sim 1.2$, if the reheating temperature is as low as the minimally allowed value by BBN, $T_R \sim 10^{-2}$~GeV, they can also account for the NANOGrav results. When we assume a different value for $r$, the values of $n_T$ and $T_R$ required to explain the NANOGrav signal are shifted, but the tendency is the same as seen by comparing the left and right panels of Fig.~\ref{fig:gamma_A}. 

\begin{figure}
\begin{center}
\includegraphics[width=\textwidth]{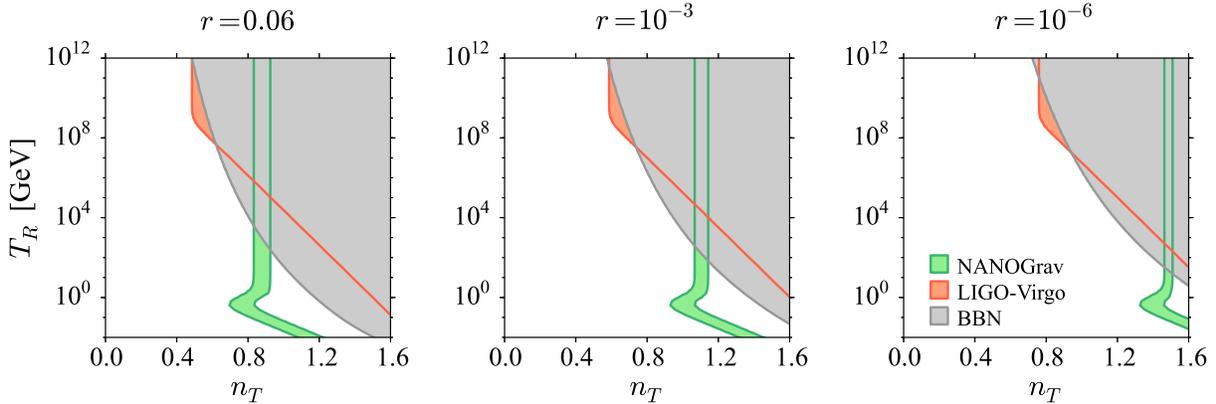}
\caption{
Parameter space consistent with the NANOGrav results (green colored region) is shown in the $n_T$--$T_R$ plane for the standard reheating scenario. The red and light gray colored regions are excluded by the current constraints from LIGO-Virgo and BBN, respectively. We show the cases for different values of the tensor-to-scalar ratio, $r=0.06, 10^{-3}, 10^{-6}$.  \label{fig:standard_nT_TR} }
\end{center}
\end{figure}

However, as already mentioned in the introduction, when we consider a blue-tilted tensor spectrum, we should also take into account the upper bound on the amplitude of the SGWB by BBN and LIGO-Virgo.  In Fig.~\ref{fig:standard_nT_TR}, we show the parameter space in the $n_T - T_R$ plane consistent with the NANOGrav results (i.e., parameter values which give prediction of $\gamma_{\rm CP}$ and $A_{\rm CP}$ falling inside the 2$\sigma$ posterior contour shown in Fig.~\ref{fig:gamma_A}) with green color for the cases of $r=0.06$ (left), $10^{-3}$ (middle), and $10^{-6}$ (right). Regions excluded by BBN and LIGO-Virgo are also depicted with light gray and red colors, respectively. One can see that a large region of the parameter space is indeed excluded by the BBN and LIGO-Virgo bounds, which typically is seen for high reheating temperature since the spectral amplitude continues to grow towards high frequencies. However, the figure illustrates the main message of this paper that, when one assumes relatively low reheating temperature (e.g. $T_R<10^3$~GeV for $r=0.06$), the spectral amplitude gets suppressed at high frequencies and IGWs can still account for NANOGrav signal without conflicting with the BBN and LIGO-Virgo constraints.

In the figure, we find that a higher value of $n_T$ is required for $T_R<1$~GeV. This is because the suppression of the GW amplitude by the MD phase occurs at the pulsar timing frequency when $T_R<1$~GeV, and a higher value of $n_T$ is needed to amplify GWs for explaining the NANOGrav results. Different panels show the cases of different values of the tensor-to-scalar ratio. For a smaller value of $r$, we find the tendency that it requires a larger value of $n_T$ to explain the NANOGrav results in order to compensate for the overall suppression of the spectral amplitude. In this case, a lower reheating temperature is required to be consistent with the BBN bound, since the spectrum grows more quickly towards high frequencies when $n_T$ is larger. 

\subsection{Case with late-time entropy production  \label{sec:entropy} }

\begin{figure}
\begin{center}
\includegraphics[width=\textwidth]{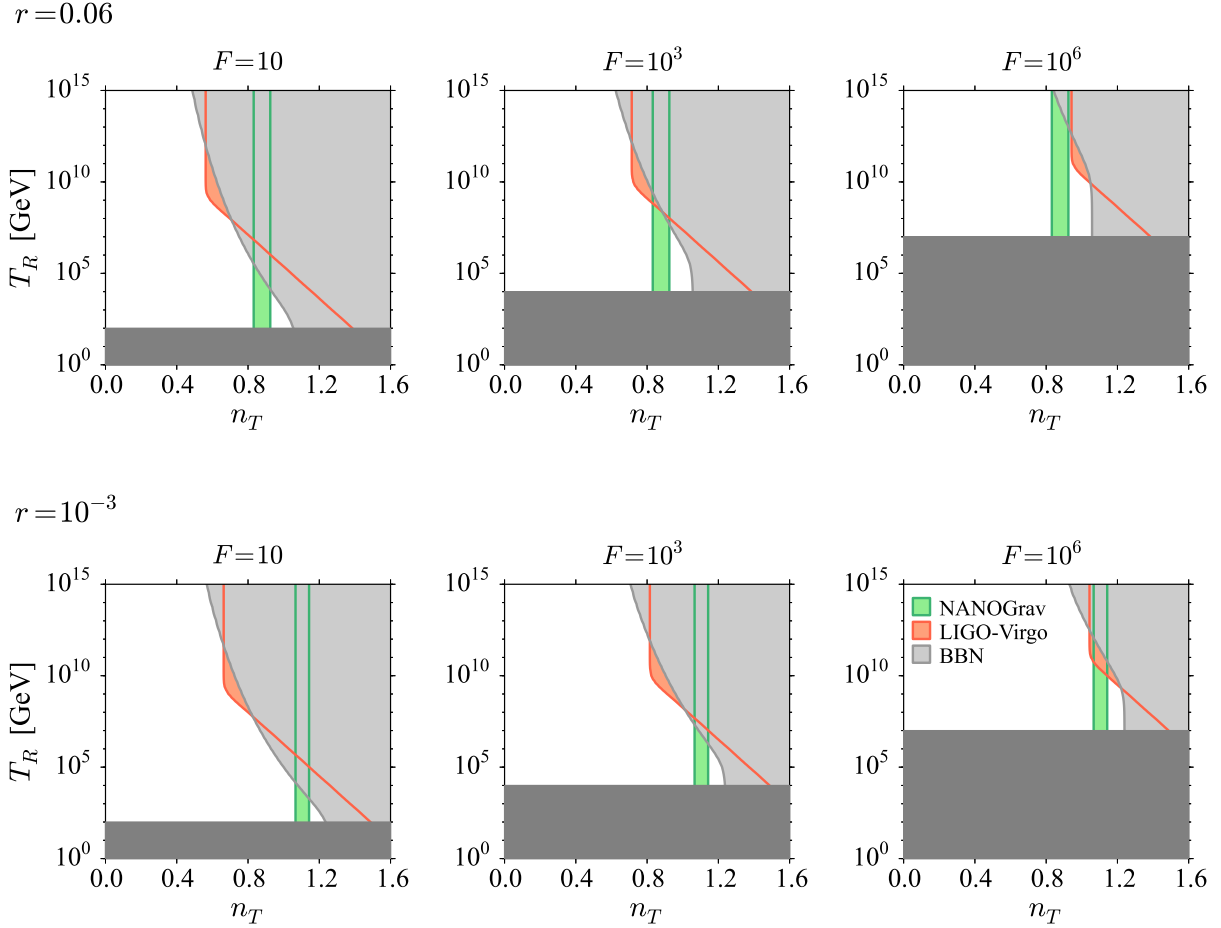}
\caption{ 
Parameter space consistent with the NANOGrav results (green colored region) is shown in the $n_T$--$T_R$ plane for the late-time entropy production scenario. The red and light gray colored regions are excluded by the current constraints from LIGO-Virgo and BBN, respectively. The second reheating temperature is set as $T_\sigma = 10$~GeV and we show the cases for different parameter sets; $F= 10$ (left), $10^3$ (middle), $10^6$ (right) and $r=0.06$ (upper), $10^{-3}$ (lower). The region covered by dark gray is not allowed because reheating temperature should be larger than the temperature of the Universe at the onset of $\sigma$-oscillation dominated epoch (see the text for details).
\label{fig:late_time_entropy_nT_TR} }
\end{center}
\end{figure}

Let us turn to the case with the late-time entropy production scenario. Since the Universe experiences MD phase twice before the BBN epoch in this case, the spectrum of IGWs gets more suppression as shown in the right panel of Fig.~\ref{fig:GW_spectra}. Therefore the BBN and LIGO-Virgo constraints become less stringent compared to the case with the standard reheating scenario. In Fig.~\ref{fig:late_time_entropy_nT_TR}, we show the parameter space consistent with the NANOGrav results as well as the excluded regions by BBN and LIGO-Virgo in the $n_T$--$T_R$ plane. The second reheating temperature is set as $T_\sigma = 10$~GeV. Different panels show the cases of different parameter values $r=0.06, 10^{-3}$ and $F= 10, 10^3, 10^6$ . 

As expected, the parameter space which can explain the NANOGrav results being consistent with the BBN and LIGO-Virgo constraints is broadened. We find that, for $r=0.06$, the reheating temperature $T_R$ can be as high as $T_R =10^8~{\rm GeV}$ for $F=10^3$. Interestingly, much higher reheating temperature, even $T_R \sim 10^{15}~{\rm GeV}$ which is close to the reheating temperature of instant reheating, is allowed if $F=10^6$. 

The dark gray region represents the requirement that the reheating temperature $T_R$ should be higher than the temperature when the $\sigma$-oscillation dominated epoch starts, $T_{\sigma R}$. Assuming entropy conservation from $T_R$ to $T_{\sigma R}$, we can replace $s(T_R)a^3(T_R)$ with $s(T_{\sigma R})a^3(T_{\sigma R})$ in the definition of $F$, Eq.~\eqref{eq:F}. Using $s\propto T^3$ and $a^3\propto \rho^{-1}\propto T^{-4}$ during the entropy production phase, we find that the requirement is 
\begin{equation}
T_{R} > T_{\sigma R} = T_\sigma F \,.
\end{equation}

\section{Conclusion \label{sec:conclusion}}

In this paper, we have investigated the implications of the recent NANOGrav results for IGWs with a blue-tilted primordial spectrum by taking into account the thermal history of the Universe after inflation. We first investigated the case of the standard reheating scenario where the MD (inflaton-oscillation dominated) phase lasts for a certain period after inflation. If the MD phase lasts long enough, the GW spectrum gets suppressed at high frequencies, and hence we can avoid the BBN and LIGO-Virgo constraints. We have shown that IGWs can explain the NANOGrav results without conflicting with the BBN and LIGO-Virgo constraints when the reheating temperature is relatively low, which can be read off from Fig.~\ref{fig:standard_nT_TR}. We found that, for $r=0.06$, $T_R\lesssim 10^3$~GeV is required to avoid the BBN constraint. We need slightly lower reheating temperature for lower value of the tensor to scalar ratio; $T_R\lesssim 10^2$~GeV for $r=10^{-3}$ and $T_R\lesssim 10$~GeV for $r=10^{-6}$.

We further considered thermal history with late-time entropy production. In this scenario, MD phases appear twice between the end of inflation and BBN epoch, and hence the amplitude of IGWs at higher frequencies gets more suppressed than the standard reheating case, which makes the constraints from BBN and LIGO-Virgo less stringent. As shown in Fig.~\ref{fig:late_time_entropy_nT_TR}, the parameter space consistent with the NANOGrav results and allowed by BBN and LIGO-Virgo is enlarged compared to the standard reheating case. In particular, when $F=10^6$ with $T_\sigma =10$~GeV and $r=0.06$, IGWs can account for the NANOGrav signal even for instant reheating, which corresponds to $T_R\sim 10^{15}$~GeV. 

Although detection of quadrupole correlations is essential to claim the discovery of the SGWB by pulsar timing arrays and further observation is needed, once the NANOGrav signal is confirmed, our result would help to test inflationary cosmology via GWs and advance our understanding of the early universe. Beyond pulsar timing, measurement of a SGWB at different frequencies by CMB B-mode polarization and future interferometer experiments (e.g. LISA, DECIGO, and Einstein Telescope) would also be crucial for testing IGWs with a blue-tilted spectrum. Such multi-frequency band observations of GWs would open up a new era in GW cosmology.

\section*{Acknowledgements}
This work is supported by JSPS KAKENHI Grant Numbers 20H01899 (SK), 20J40022 (SK), 17H01131 (TT), 19K03874 (TT), 20H01932 (SY), 20K03968 (SY) and MEXT KAKENHI Grant Number 19H05110 (TT). SK is supported by the Atracción de Talento contract no. 2019-T1/TIC-13177 granted by the Comunidad de Madrid in Spain.

\clearpage 

\bibliography{NANOGrav_inflation}

\end{document}